\documentclass{article}

\usepackage{arxiv}

\usepackage[utf8]{inputenc} %
\usepackage[T1]{fontenc}    %
\usepackage{hyperref}       %
\usepackage{url}            %
\usepackage{booktabs}       %
\usepackage{amsfonts}       %
\usepackage{nicefrac}       %
\usepackage{microtype}      %
\usepackage{lipsum}		%
\usepackage{epstopdf}
\usepackage{doi}
\usepackage{parskip}
\usepackage{color}
\usepackage{caption}
\usepackage{subcaption}
\usepackage{float}
\usepackage{mathematics}
\usepackage{graphics,graphicx}
\usepackage{wrapfig}
\usepackage{amsmath}
\usepackage{cancel}
\usepackage{mathabx}
\usepackage{enumitem}
\usepackage{cite} 
\usepackage{todonotes}
\usepackage{soul}
\usepackage{dsfont}
\usepackage{verbatim}
\usepackage{appendix}
\usepackage{multirow}
\usepackage{upgreek}

\newcommand\includegraphicsifexists[2][width=\linewidth]{\IfFileExists{#2}{\includegraphics[#1]{#2}}{}}

\newcommand{\Ell}{\mathcal{L}}

\newcommand{\C}{\mathds{C}}

\newcommand{\Tant}{\emph{Dev1}}
\newcommand{\Geb}{\emph{Dev2}}

\title{Data-Driven Characterization of Latent Dynamics on Quantum Testbeds}

\author{ 
\href{https://orcid.org/0000-0001-6882-9737}{\includegraphics[scale=0.06]{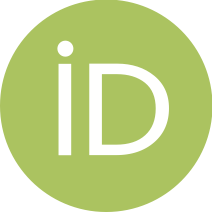}\hspace{1mm}
    Sohail Reddy}\thanks{Corresponding author.} \\
	Lawrence Livermore National Laboratory \\
	Livermore, CA 94550 \\
	\texttt{reddy6@llnl.gov} \\
	\And
	\href{https://orcid.org/0000-0000-0000-0000}{\includegraphics[scale=0.06]{Figures/orcid.png}\hspace{1mm}
	Stefanie Günther} \\
	Lawrence Livermore National Laboratory \\
	Livermore, CA 94550 \\
	\texttt{guenther5@llnl.gov} \\
	\And
	\href{https://orcid.org/0000-0002-2437-7844}{\includegraphics[scale=0.06]{Figures/orcid.png}\hspace{1mm}
	Yujin Cho} \\
	Lawrence Livermore National Laboratory \\
	Livermore, CA 94550 \\
	\texttt{cho25@llnl.gov} \\
}

\renewcommand{\shorttitle}{Data-Driven Characterization of Latent Dynamics on Quantum Testbeds}
\newcommand{\etal}{\textit{et al.}}

\begin{document}
\maketitle

\begin{abstract}
This paper presents a data-driven approach to learn latent dynamics in superconducting quantum computing hardware. To this end, we augment the dynamical equation of quantum systems described by the Lindblad master equation with a parameterized source term that is trained from experimental data to capture unknown system dynamics, such as environmental interactions and system noise. We consider a structure preserving augmentation that learns and distinguishes unitary from dissipative latent dynamics parameterized by a basis of linear operators, as well as an augmentation given by a nonlinear feed-forward neural network. Numerical results are presented using data from two different quantum processing units (QPU) at Lawrence Livermore National Laboratory's Quantum Device and Integration Testbed. We demonstrate that our interpretable, structure preserving, and nonlinear models are able to improve the prediction accuracy of the Lindblad master equation and accurately model the latent dynamics of the QPUs.
\end{abstract}


\section{Introduction} \label{sec:Intro}

As the size of quantum systems increases rapidly to be able to perform complex scientific simultaions, focus has now shifted to quantum error correction and demonstrations of practical applications of quantum computing. Any such application requires accurate control of a Noisy Intermediate-scale Quantum (NISQ) system that is plagued with sensitivity to a myriad of noise sources \cite{preskill2018quantum}. Computing these control pulses requires knowledge of the underlying quantum mechanical model governing the quantum system. While quantum characterization aims to determine the device parameters that define the numerical model to simulate the quantum dynamics, determining the appropriate forms and parameters of quantum master equations for real systems is challenging and typically requires continuous parameter estimation or discrete operator tomography to fit unknown parameters experimentally~\cite{cubitt2012complexity}. Deducing model parameters such as transition frequencies and decoherence times requires designing specific protocols, which become increasingly difficult for larger systems. The resulting parameters are \textit{estimates} of the true underlying device frequencies, and combined with the shortcomings of the quantum dynamical model (i.e. Lindblad master equation) itself, yield only an \textit{approximate} prediction of the underlying dynamical processes where the modeling errors propagate directly to numerically optimized control pulses. More accurate models will yield more robust control pulses, help identify latent noise processes which can greatly inform hardware development and serve as efficient emulators of the quantum devices. 

In order to reduce the modeling errors, we consider a data-driven approach to quantum characterization that augments an approximate dynamical model of the quantum device by an additional source that is trained from measurement data to capture latent quantum dynamics. The trained source term represents a \textit{correction} that accounts for any unknown dynamical processes arising from many sources: latent environmental interactions and noise processes, drift in system frequencies or other Hamiltonian parameters, losses along the control lines of the device, or qubit cross talk. We include the learnable parameters directly into the Lindblad's master equation that describes the known underlying dynamics, rather than modeling the entire dynamical system by a machine learning model. This augmented model describes a Universal Differential Equation (UDE) \cite{rackauckas2020universal}, also termed physics-informed Neural Ordinary Differential Equation \cite{lai2021structural}. UDE employs a machine learning agent as a correction to an underlying (approximate) physical model described by a differential equation that stems from possible prior knowledge of a dynamical system. 
A similar framework has successfully been applied to learn high-dimensional structural dynamical systems \cite{lai2021structural}, 
general nonlinear coupled oscillators \cite{koch2023structural}, and to predict back hole dynamics \cite{keith2021learning}. 
Those applications have demonstrated that such a `gray-box' learning approach, which combines black-box machine learning models with a white-box approach that approximately models known dynamics yields, good generalization properties, is resilient to noisy data, and requires a smaller training data set.  
On the other hand, integrating physical knowledge into the learning approach for dynamical systems has most prominently been done via physics-informed neural networks (PINNs) \cite{cuomo2022scientific, faroughi2022physics, huang2022partial,kim2021knowledge}. These approaches typically train a machine learning agent to predict the state of the dynamical system directly, while enforcing the underlying physics by embedding the residual of the governing equations in the loss function. However, these methods typically assume that the exact governing equations are known. 

In the quantum domain, machine learning approaches have recently been applied to a variety of applications \cite{gebhart2023learning, dunjko2018machine, carleo2019machine} including Hamiltonian estimation \cite{gentile2021learning, flynn2022quantum, Wang:2017}, quantum phase estimation \cite{nautrup2019optimizing}, quantum error correction \cite{fosel2018reinforcement, nautrup2019optimizing}, quantum control \cite{bukov2018reinforcement, schafer2021control, schafer2021control}, and to learn state discrimination for multi-qubit readout from data \cite{lienhard2022deep}. To predict and accelerate quantum dynamical calculations, Rodríguez \etal \cite{rodriguez2022comparative} provides a benchmark study of 22 supervised machine learning methods and compare their ability to forecast long-time dynamics. The vast majority of these methods can be considered as black-box approaches, where a machine learning model (primarily, neural networks) approximates the input-output relation of quantum dynamical processes directly \cite{mohseni2023deep, zhu2023predictive, huang2022learning, lewis2023improved}. Deep quantum neural-networks have previously been used to solve Lindblad master equations for open quantum many-body systems \cite{liu2022solving, hartmann2019neural} and models governing non-Markovian dynamics \cite{cerrillo2014non}. Recurrent neural networks (RNNs) were used by Leclerc \cite{leclerc2021predicting} to generate an input-output relationship of dissipative mechanisms to time evolution of measurement outcomes. They have also been applied to predict quantum trajectories for transmon qubits under different microwave control sequences \cite{flurin2020using}. While these studies use RNNs to model the input-output relation directly, time-derivative learning using Neural ODEs has been demonstrated on synthetic problems to learn corrections to the Hamiltonian as well as a recurrent non-Markovian Lindblad operators\cite{banchi2018modelling}. Such Neural ODEs have also been extended to encode completely-positive trace-preserving (CPTP) maps by rewriting the system Hamiltonian, collapse operators, and non-Markovian memory kernels as neural-net-like expressions (RNN layers), demonstrating that a comparatively smaller training data set is required to achieve accurate model fidelity \cite{krastanov2020unboxing}. Other gray-box approaches that combine black-box neural networks with numerically evolved quantum dynamical processes have demonstrated superior performance over model fitting learning approaches~\cite{youssry2022experimental}, and allow for better interpretability; see for example for the extraction of noise power spectra \cite{youssry2020characterization} and \cite{luchnikov2022probing} for automatic extraction of the effective environment dimensionality as well as eigenfrequencies of the joint system and environment dynamics.  

This work considers such a `gray-box' approach where we use the Lindblad's master equation as a baseline dynamical model and augment it with source terms that are trained using measured data. We investigate a structure-preserving approach that aims to distinguish between unitary and dissipative dynamics, and allow for interpretability of the learned model. We also consider an augmentation by a neural network model to capture any nonlinear interaction that might be present. We apply the approach to identify latent dynamics of two superconducting quantum processing units (QPUs) in the Quantum Design and Integration Testbed (QuDIT) at Lawrence Livermore National Laboratory (LLNL).

\section{Baseline Model of Open Quantum Systems: Lindblad Master Equation} \label{subsec:MathModel}

We consider the Lindblad master equation as an approximate underlying baseline model for open quantum system dynamics, 
\begin{equation} \label{eq:MasterEq}
  \dfrac{\partial \rho}{\partial t} = -i \left[ H, \rho \right] + \Ell(\rho),
\end{equation}
where $\rho \in \C^{N\times N}$ is the density matrix, $[\cdot,\cdot]$ is the commutator, $H \in \C^{N\times N}$ is the Hamiltonian governing unitary dynamics and $\Ell(\rho)$ is a Lindbladian operator that models decoherence. We consider single qubit dynamics, hence $N=2$, however the approach is straightforward to generalize to multiple qubits. When $\Ell(\rho)=0$, Eq. \eqref{eq:MasterEq} is called the Liouville-von Neumann (LvN) equation and represents the dynamics of a closed quantum system in density matrix formalism. We target superconducting quantum devices and model the Hamiltonian in the rotating frame as
\begin{equation} \label{eq:Hamiltonian}
  H(t) := \left( \omega - \omega^{rot} \right) a^\dagger a +  p(t) \left(a + a^\dagger \right) + iq(t) \left(a - a^\dagger\right)
\end{equation}
where $\omega$ denotes the qubit transition frequency, $\omega^{rot}$ denotes the frequency of rotation in the rotating wave approximation (RWA), and $a$ and $a^\dagger$ denote the lowering and raising operators, respectively. The rotating-frame control functions $p(t): \R \mapsto \R$ and $q(t):\R \mapsto \R$ denote the in-phase and quadrature components of the microwave pulses that drive the quantum device. The control pulse in the laboratory frame is then $f(t) = (p(t)+iq(t))e^{i\omega^{rot}}$.
The Lindblad operator is of the form
\begin{equation} \label{eq:Lindbladian}
    \Ell(\rho) := \sum_{i=1}^{N_\Ell} \tau_i \rbrac{\Ell_i \rho \Ell_i^\dagger - \dfrac{1}{2} \left\lbrace \Ell_i^\dagger \Ell_i , \rho \right\rbrace}
\end{equation}
where $N_\Ell$ are the total number of decoherence operators $\Ell_i$ (also called jump operators), and $\left\lbrace \cdot , \cdot \right\rbrace$ is the anti-commutator. We here consider $N_{\Ell}=2$ standard Lindblad operators in our baseline model, namely $\Ell_1 := a$ and $\Ell_2 := a^\dagger a$, where $\tau_1 = 1/T_1$ and $\tau_2 = 1/T_2$ are the energy decay and dephasing rates, respectively.

\section{Data-Driven Modeling Through Universal Differential Equations} \label{sec:UDE}

In the subsequent sections, we introduce a parameterized trainable source term $\mathcal{S}(\rho)$ that acts as a correction to the dynamical equation Eq. \eqref{eq:MasterEq} and that is trained from data:
\begin{equation} \label{eq:MasterEqUDE}
  \dfrac{\partial \rho}{\partial t} = -i \left[ H, \rho \right] + \Ell(\rho) + \mathcal{S}(\rho).
\end{equation}
The form of this source term is introduced below. We refer to the source-free models as \emph{base} models which captures already identified (approximate) dynamics. Together with the trained source term, Eq. \eqref{eq:MasterEqUDE} then constitutes a universal differential equation, combining known and approximate dynamics with a data-driven correction term.

\subsection{Learning Structure Preserving Operators} \label{subsec:SPLearning}

We first consider a structure preserving ansatz for generating and distinguishing unitary from dissipative dynamics. First, we consider a splitting of the learnable source term $S(\rho)$ of the form
\begin{equation} \label{eq:SPAnsatz}
\begin{aligned}	
	\dfrac{\partial \rho}{\partial t} = -i \left[ H + \mathcal{S}_H, \rho \right] + \Ell(\rho) + \mathcal{S}_\Ell(\rho) 
\end{aligned}  
\end{equation}  
where $\mathcal{S}_H$ is a Hermitian operator that corrects for latent unitary dynamics, whereas $\mathcal{S}_{\mathcal{L}}(\rho)$ is constructed to correct for latent dissipative dynamics.
We parameterize the generator for unitary dynamics $\mathcal{S}_H$ as a linear combination of the generalized Gell-Mann matrices, $\left\lbrace \Lambda_1 ,\ldots, \Lambda_{N^2-1} \right\rbrace$, spanning the Lie algebra of $\mathrm{SU(N)}$
\begin{equation} \label{eq:HermitianAnsatz}
	\mathcal{S}_H := \sum_{j=1}^{N^2 - 1} \alpha_j \rbrac{ \Lambda_j - \bra{0}\Lambda_j\ket{0} I},
\end{equation}
where the coefficients $\alpha_j \in \R$ are trainable parameters and the second term in the summation shifts the energy spectrum such that the ground state energy is 0; i.e. $\bra{0} (H + \mathcal{S}_H) \ket{0} = 0$. Note that for $N=2$, $\Lambda_i$, where $i=1,2,3$, correspond to the Pauli matrices with $\Lambda_1 := \sigma_x$, $\Lambda_2 := \sigma_y$ and $\Lambda_3 := \sigma_z$.

To parameterize the learnable dissipative source term, we consider a generalized form of the Lindblad operator, following approprate diagonalization by unitary transformation of the coefficient matrix (see \cite{Breuer2007}):
\begin{equation} \label{eq:LindbladianAnsatz}
	\mathcal{S}_\Ell(\rho) := \sum_{j=1}^{N^2-1} \gamma_j \underbrace{\left( \widecheck{\Lambda}_j \rho \widecheck{\Lambda}_j^\dagger - \dfrac{1}{2} \left\lbrace \widecheck{\Lambda}_j^\dagger \widecheck{\Lambda}_j , \rho \right\rbrace \right)}_{\mathcal{S}_\Ell(\rho;\widecheck{\Lambda})}
\end{equation}
where we take the collapse/jump operators $\widecheck{\Lambda}_j$ to be the upper triangular part of the generalized Gell-Mann matrices $\Lambda_j$ and $\gamma_j\in \R$ denotes a trainable parameter. The decoherence rate for the collapse process described by a jump operator $\widecheck{\Lambda}_j$ is $\gamma_j$. Since Lindblad's master equation is invariant under unitary and inhomogenous transformation, the decomposition into the Hamiltonian and dissipative parts is not unique \cite{Manzano2020,Gegg2016,Oppenheim2023,Tarnowski2021}. Hence, an alternate basis $\widecheck{\Lambda}$ can be defined. We motivate our choice of basis by the fact that in the case of Hermitian jump operators, the purity of the system fulfills $\frac{d}{dt}(\mathrm{Tr}[\rho^2]) \leq 0$, a condition not satisfied by the experimentally obtained density matrices.
Furthermore, it is straightforward to see that $\mathcal{S}_\Ell(\rho;\widecheck{\Lambda}) = \mathcal{S}_\Ell(\rho;i \widecheck{\Lambda})$ and for $N=2$, we see that $\widecheck{\Lambda}_1 = i \widecheck{\Lambda}_2  = a$ which represents $T_1$ energy decay. Hence, $\widecheck{\Lambda}_1$ and $\widecheck{\Lambda}_3$ are sufficient to describe the dissipative dynamics for a single qubit case. Despite this, we also include $\widecheck{\Lambda}_2$ in our ansatz for ease of implementation and generalization to larger multi-level systems. For a single qubit system, the perturbation to the energy decay rate is given by $\Delta \tau_1 = \gamma_1+\gamma_2$. Furthermore, note that $\mathcal{S}_\Ell(\rho;\widecheck{\Lambda}_3) = 4 \mathcal{S}_\Ell(\rho; a^\dagger a)$ \footnote{This is easily seen by constructing the superoperator form of this jump operator.} which represents the $T_2$ dephasing processes such that the dephasing rate perturbation is $\Delta \tau_2 = 4\gamma_3$. The effective and perturbed decoherence times are then given as $T_i = (\tau_i + \Delta \tau_i)^{-1}$.

Learning the unitary and dissipative latent dynamics within this structure-preserving ansatz amounts to identifying coefficients $\alpha_i, \gamma_i \in \R, i=1,\dots, N^2-1$ that yield the highest fidelity when comparing the dynamics of the augmented model \eqref{eq:MasterEqUDE} to device data. This choice of parameterizing $\mathcal{S}_H$  and $\mathcal{S}_{\Ell}(\rho)$ ensures that \eqref{eq:SPAnsatz} is a generator for \emph{Completely-Positive Trace-Preserving (CPTP)} maps \cite{nielsen2010quantum}, and therefore, enforces physically meaningful quantum state evolution.

\subsection{Learning Neural Network Operators} \label{subsec:NNLearning}

A second, more general approach for learning quantum dynamical operators involves modeling the source term as a feed-forward neural network. In this approach, the source $\mathcal{S}(\rho)$ is an $L$-fold composition of network layers $\mathcal{N}_l$
\begin{equation} \label{eq:NNAnsatz}
	\mathcal{S}(\rho) =  \mathcal{N}(\rho) := \mathcal{N}_{L} \circ \mathcal{N}_{L-1} \circ \ldots \circ \mathcal{N}_{1} (\rho)
\end{equation}
where each layer $\mathcal{N}_l: \R^{N^2} \rightarrow \R^{N^2}$ consists of an affine transformation followed by a nonlinear activation $\sigma : \R \rightarrow \R$ that is applied component-wise:
\begin{equation} \label{eq:NN}
	\mathcal{N}_l(\mathbf{x}) = \sigma_l \left( \mathbf{W}_l \mathbf{x} + \mathbf{b}_l  \right)
\end{equation}
where $\mathbf{W}_l$ and $\mathbf{b}_l$ are the weight matrix and bias vector, respectively, for layer $l$. 
Hermiticity of the density matrix is preserved by letting $\mathcal{N}:\Vec{\alpha} \mapsto \Vec{\hat{\alpha}}$  map a density matrix to a Hermitian source term, both expanded in a Hermitian basis $\hat H$ with expansion coefficients $\Vec{\alpha} \in \R^N$ and $\Vec{\hat{\alpha}} \in \R^N$, respectively.
Although $\hat H$ can be any basis that spans the vector space of Hermitian matrices, we employ the basis defined as 
\begin{equation} \label{eq:TrivialBasis}
	\hat{H}_{\Vec{i}} = \hat{H}_{jk} =  \begin{cases}
	\ketbra{j}{j} & j=k\\[5pt]
	\dfrac{1}{2}\rbrac{\ketbra{j}{k} + \ketbra{k}{j}} & j<k \\[5pt]
	\dfrac{i}{2}\rbrac{\ketbra{j}{k} - \ketbra{k}{j}} & j>k 
	\end{cases}
\end{equation} 
where $\Vec{i} = (j,k)$ is a multi-index, as they allow easy computation of the expansion coefficients due to their trace orthonormality (i.e. $\mathrm{Tr}\left(\hat{H}_{\Vec{i}} \hat{H}_{\Vec{j}} \right) = \delta_{\Vec{ij}}$).

The trainable parameters of this ansatz are the weights, $\mathbf{W}_l$, and biases, $\mathbf{b}_l$, of each layer of the neural network.
In the absence of the nonlinear activation function (i.e. $\sigma$ is the identity), then the $L$-fold composition reduces to an affine map. In the results presented below, we will investigate both a nonlinear model using the $tanh$ activation function, as well as an affine model for the source term using the identity activation function.

Unlike the structure-preserving ansatz in Section \ref{subsec:SPLearning}, the neural-network ansatz does not guarentee a CPTP map. 
To obtain a valid density matrix, we apply a spectral filter with renormalization. In particular, we construct filtered density matrices as
\begin{align} \label{eq:SpectralFilter}
	\overline{\rho} = \sum_{i=1}^N \mathcal{E}_i | \Psi_i \rangle \langle \Psi_i | 
    \quad \text{with} \quad
	\mathcal{E}_i = \dfrac{\mathcal{H}(E_i)\cdot E_i}{\sum_j^N \mathcal{H}(E_j)\cdot E_j},
\end{align}
where $\cbrac{\Psi_i}_{i=1}^N$ and $\cbrac{E_i}_{i=1}^N$ are the eigenvectors and the corresponding unfiltered eigenvalues, respectively, $\cbrac{\mathcal{E}_i}_{i=1}^N$ are the filtered and renormalized eigenvalues, and the filter function $\mathcal{H}(x)$ is the Heaviside step-function. Due to the increased computational cost of the spectral decomposition, we do not apply the filter during the training phase, but only during the deployment phase of the trained model.

\subsection{Training Procedure} \label{subsec:Training}

The training procedure for learning the structure preserving model as well as the neural network-based source term amounts to solving an optimization that is constrained by the augmented dynamical equation \eqref{eq:MasterEqUDE}. Denote by $\boldsymbol{\xi}$, a set of experiments that are parameterized by control functions which determine the time-evolution of the true/exact quantum state. For the examples below, this set consists of various microwave control pulses, $p(t), q(t)$ in \eqref{eq:Hamiltonian} which will be described in Section \ref{sec:qudit}. Each experiment defines thetrue/exact evolution of the density matrix, $\widetilde{\rho}(t_j; \xi_i)$, which is estimated using quantum state tomography \cite{Qi:2013} at various time steps $t_j \in (0,T]$. In this work, we apply linear inversion followed by spectral filtering and renormalization, Eq. \ref{eq:SpectralFilter}. Let $\boldsymbol{\theta}$ denote the set of learnable parameters in the augmented differential equation, i.e. $\boldsymbol{\theta} = \left\lbrace \alpha_l \bigcup \gamma_l: l=1,\ldots,N^2-1 \right\rbrace$ for the structure preserving ansatz in Section \ref{subsec:SPLearning} and $\boldsymbol{\theta} = \left\lbrace \mathbf{W}_l \bigcup \mathbf{b}_l : l = 1,\ldots, L \right\rbrace$ for the neural network model defined in Section \ref{subsec:NNLearning}. The goal of the training procedure is then to find optimal parameters $\boldsymbol{\theta}$ such that the resulting augmented differential equation matches the measured data $\widetilde{\rho}$ at time-steps $t_j$. To this end, the following constraint optimization problem is solved: 
\begin{subequations} \label{eq:OptimizationProb}
	\begin{align}
		\begin{split} \label{eq:LossFunction}
			\boldsymbol{\theta}^\star := & \arg\min_{\boldsymbol{\theta} \in \boldsymbol{\Theta}} ~~ \sum_{i=1}^{|\boldsymbol{\xi}|}  \sum_{j=1}^{N_T} \lVert \widetilde{\rho}(t_j;\xi_i) - \rho(t_j;\xi_i,\boldsymbol{\theta})\rVert^2_F
		\end{split}\\
		\begin{split} \label{eq:Constraint}
			\mathrm{subject~to~} &: \dfrac{\partial \rho}{\partial t} = -i \left[ H, \rho \right] + \Ell(\rho) + \mathcal{S}(\rho;\boldsymbol{\theta}), \ \forall t \in (0,T] 
		\end{split}
	\end{align}	
\end{subequations}
In order to evaluate the objective function, the augmented dynamical system \eqref{eq:Constraint} is evolved numerically for the given experimental setup in $\boldsymbol{\xi}$ and the current parameters $\Vec{\theta}$. Although any other numerical solver can be used, we here use an explicit 4th-order Runge-Kutta scheme for the numerical time-integration of the dynamical system. This scheme yields the predicted evolution $\cbrac{\rho(t_j)}_{j=1}^{N_T}$ which is compared to the true evolution within the loss function in terms of their Frobenius distance. 
We utilize the Julia SciML \cite{rackauckas2017differentialequations} framework to solve this optimization problem with the ADAM optimizer mini-batches as an initial optimizer, followed by a full-batch L-BFGS optimization. The gradient is calculated using automatic differentiation. 

After training on a defined set of experiments and time-horizon $0<t_j<T$, we investigate the efficacy of the trained augmented model on a set of validation experiments. In particular, we show that the trained model is able to accurately predict the time-evolution on time domains much longer than the training domain, for experimental setups that have not been included in training. 
We refer to the accuracy over the training and validation sets as \emph{interpolation} and \emph{extrapolation} accuracy, respectively.

In the sections that follow, we will compare the solutions obtained using different formulations using the trace distance $T(\widetilde{\rho},\rho) = \frac{1}{2}\mathrm{Tr}\left[ \sqrt{(\widetilde{\rho}-\rho)^\dagger (\widetilde{\rho}-\rho)} \right]$. %
In each case, we will apply the spectral filter in Eq. \ref{eq:SpectralFilter} to obtain a valid density matrix. To quantify the generalization (i.e. extrapolation) properties of the trained models, we employ the expected trace distance
\begin{equation} \label{eq:ExpTraceDist}
    \mathbb{E} \left[ T \left(\widetilde{\rho},\rho(\boldsymbol{\theta}) \right) \right] = \int_{\Omega_{\Vec{\xi}}} T \left(\widetilde{\rho}(\boldsymbol{\xi}),\rho(\boldsymbol{\xi;\theta}) \right) \pi(\boldsymbol{\xi}) d\boldsymbol{\xi}
\end{equation}
where $\pi(\boldsymbol{\xi})$ is a multivariate uniform distribution over the parameterized experiments. We estimate Eq. \ref{eq:ExpTraceDist} using Monte Carlo sampling.

\section{Learning Latent Dynamics of LLNL Testbed's QPUs} \label{sec:qudit}

We apply the learning strategy to identify the latent dynamics of two quantum processing units (QPUs), \Tant~and \Geb, at LLNL's Quantum Design and Integration Testbed (QuDIT). \Tant~is a 2D single transmon made of Tantalum on a sapphire substrate \cite{Place2021}. \Geb~is a single 3D transmon mounted in a high-purity aluminum resonator. Both QPUs are mounted at 10 mK in a dilution fridge.
The parameters of the base models, such as transition frequencies ($\omega_{01}$), energy decay times ($T_1$), and dephasing times ($T_2$) were estimated using standard characterization protocols, using standard Rabi, Ramsey, and energy decay measurements. The measured parameters are given in Tab. \ref{table:QPUParams}. We apply our data-driven technique on both QPUs to investigate its performance on devices that exhibit different noise levels.
\begin{table}[h] %
\caption{Parameters of two QPUs at LLNL's Quantum Design and Integration Testbed. $\omega_{01}$ indicates the qubit transition frequency between $|0\rangle$ and $|1\rangle$. $T_1$ was measured via standard energy-decay experiment and $T_2$ via Ramsey experiments. \Tant~exhibits much longer coherence times than \Geb.} \label{table:QPUParams}
\centering
\begin{tabular}{c | c c c } 
 \hline
 QPU & $\omega_{01}$ (GHz) & $T_1$ ($\upmu$s)& $T_2$ ($\upmu$s) \\
 \hline 
 \Tant & 3.448 & 214 & 32 \\
 \Geb & 4.086 & 62 & 6 \\
 \hline
\end{tabular}
\end{table}

\begin{figure}[h] \centering
\begin{subfigure}[h]{0.45\textwidth} %
\includegraphicsifexists[width=\textwidth]{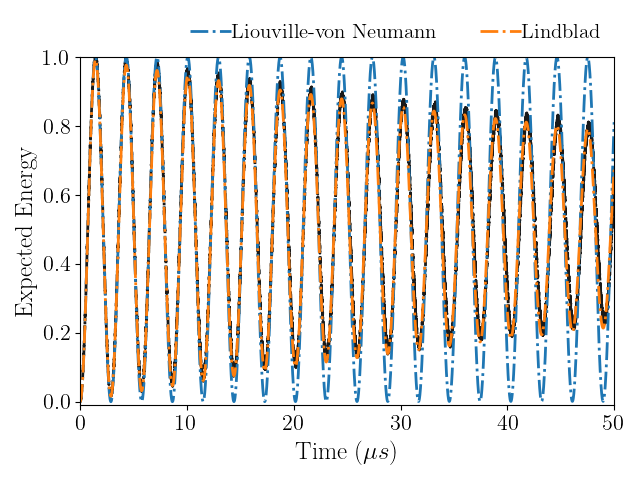}  \label{fig:Tant:Base:ExpEnergy} 
\end{subfigure}
\begin{subfigure}[h]{0.45\textwidth} %
\includegraphicsifexists[width=\textwidth]{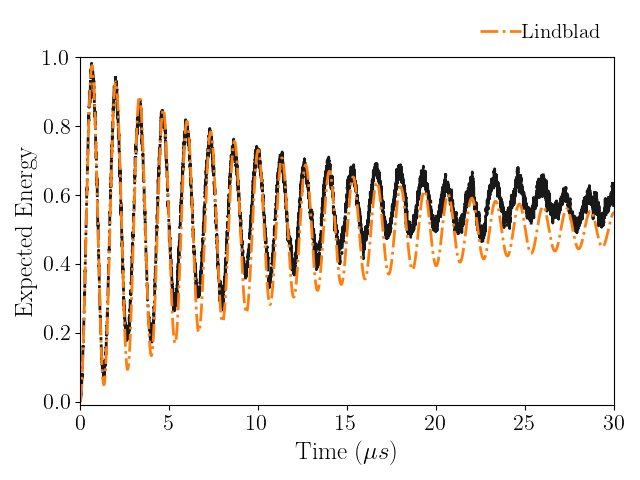}  \label{fig:Geb:Base:ExpEnergy} 
\end{subfigure}
\begin{subfigure}[h]{0.45\textwidth} %
\includegraphicsifexists[width=\textwidth]{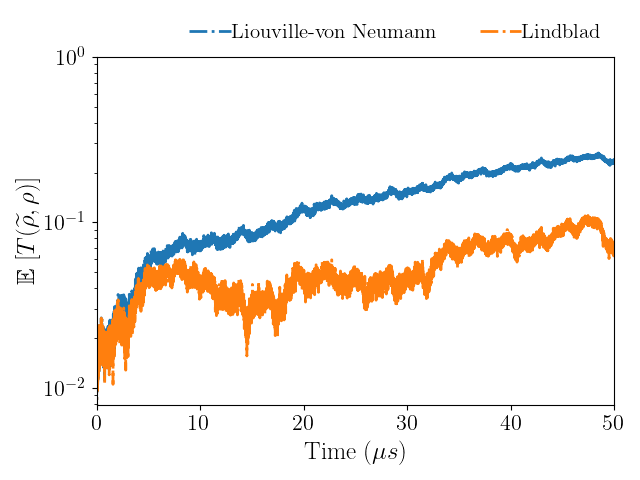}  \label{fig:Tant:Base:TrDist} 
\end{subfigure}
\begin{subfigure}[h]{0.45\textwidth} %
\includegraphicsifexists[width=\textwidth]{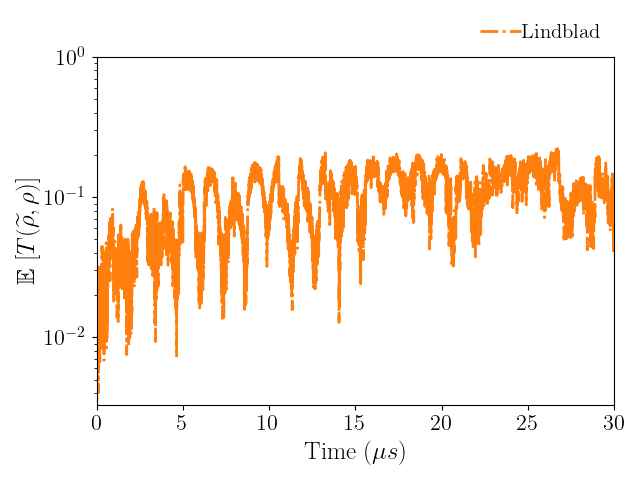}  \label{fig:Geb:Base:TrDist} 
\end{subfigure}
\captionsetup{singlelinecheck=off,font=footnotesize}
\caption[]{Time evolution of the expected energy of a sample experiment (top) and expected trace distance (bottom) on \Tant ~(left) and \Geb ~(right) compared to the underlying base models. Black line represents experimentally measured data.}
\label{fig:Base:ExpEnergy}
\end{figure}

To generate the training and validation data sets, we apply constant square pulses in the rotating frame at $\omega_{01}$. The set of experiments $\Vec{\xi}$ consists of various pulse amplitudes $\xi_j \sim \mathcal{U}(0,p_{max})$ with a maximum amplitude of $p_{max} = 3.47$~MHz for \Tant~and $p_{max} = 1.25$~MHz for \Geb. The pulses drive the qubits between the ground and first excited state (see Fig. \ref{fig:Base:ExpEnergy}). The pulses are applied for a total duration of $T=50~\upmu$s and $T=30~\upmu$s on \Tant~and \Geb, respectively, with measurements taken every $4~$ns. A total of 5000 shots are performed and the measured states are classified using a Gaussian Mixture Model (GMM). The density matrices at each timestep are then estimated by quantum state tomography using linear inversion estimate (LIE)~\cite{Qi:2013}.
The resulting Bloch vector is projected onto $x$, $y$, and $z$ basis, which yields the population vector $\Vec{p}$.
The vectorized density matrix, $\mathrm{vec(\rho)}$, can then be obtained as
\begin{align}
    \mathrm{vec(\rho)} = \Vec{M}^{-1}\Vec{p} 
    \quad \text{with} \quad
    \Vec{p}=\begin{pmatrix}1\\2P(x)-1\\2P(y)-1\\2P(z)-1\end{pmatrix},\quad 
    \Vec{M}=
    \begin{pmatrix}
    1 & 0 & 0 & 1 \\
    0 & -1 & -1 & 0 \\
    0 & i & -i & 0 \\
    -1 & 0 & 0 & 1
    \end{pmatrix}
\end{align}

Figure \ref{fig:Base:ExpEnergy} shows the time-evolution of the expected energy of the density matrices obtained through LIE for each device for a sample experiment, as well as the predicted evolution from the underlying base models in terms of Lindblad's and Liouville-von-Neumann (LvN) equation.
For \Tant, we see a good agreement between the experimental data and the Lindblad model; the LvN model shows comparable accuracy for short durations ($t \leq 6~\upmu$s) before the dissipative effects begin to dominate. We employ both the Lindblad and LvN equations as the base models when learning this QPU's dynamics. In the case of \Geb, we see that the Lindblad model fails to accurately predict the population evolution over time. This suggests that the Lindblad model described by the Hamiltonian in Eq. \ref{eq:Hamiltonian}, and $T_1$ and $T_2$ decoherence operators, does not capture all of the processes present on the testbed. Therefore, we only consider the Lindblad model as the baseline model when learning \Geb's dynamics.

\subsection{Learning Dynamics of QPU - \Tant} \label{subsubsec:qudit:tantalum}

We first investigate our UDE approach for learning dynamics of \Tant~QPU that has longer coherence times. To investigate the effects of size of the training dataset, we train the models with data over two different durations, $T_{Tr.} = 10~\upmu$s and $T_{Tr.} = 20~\upmu$s; the remaining data over $T_{Tr.} < t \leq 50~\upmu$s is used for validation.  Here we learn a single source term $\mathcal{S}(\rho)$ using training data from a total of five experiments (i.e. $|\Vec{\xi}|=5$ in Eq. \ref{eq:OptimizationProb}), to learn an operator that is control-independent. We refer to such an operator as an \emph{Experiment-Generalized} operator. The results obtained using the structure preserving ansatz will be denoted by $\mathcal{SP}$ and those obtained using the affine and nonlinear models will be denoted by $\mathcal{N}_A$ and $\mathcal{N}_N$, respectively.

\begin{figure}[h] \centering
\begin{subfigure}[h]{\textwidth} %
\centering
\includegraphicsifexists[width=0.45\textwidth]{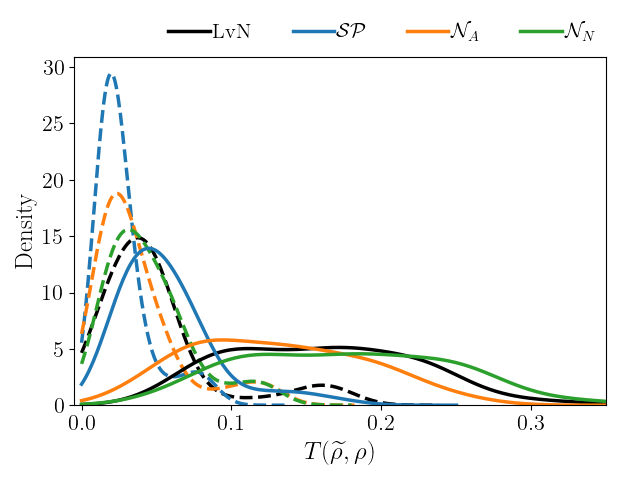} %
\includegraphicsifexists[width=0.45\textwidth]{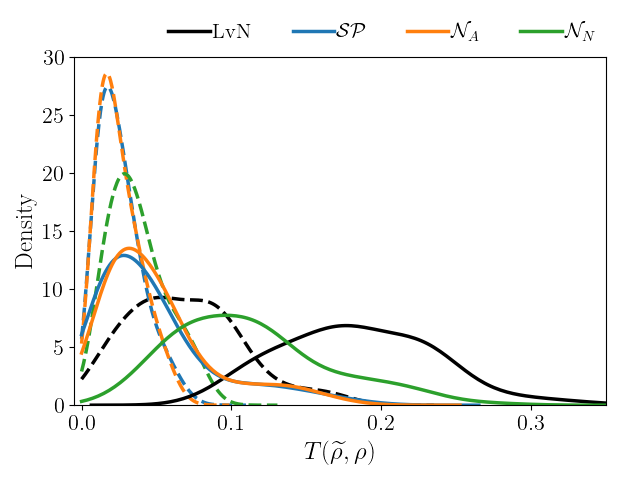} %
\label{fig:Tant:HBase:TrDist} 
\caption{Base model: Liouville von-Neumann}
\end{subfigure}
\begin{subfigure}[h]{\textwidth} %
\centering    
\includegraphicsifexists[width=0.45\textwidth]{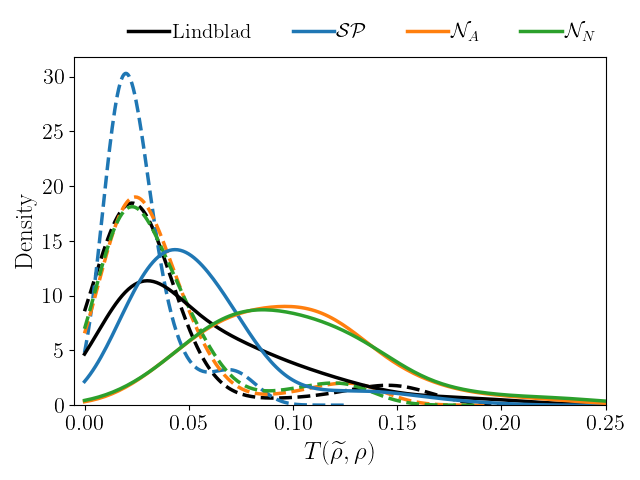} %
\includegraphicsifexists[width=0.45\textwidth]{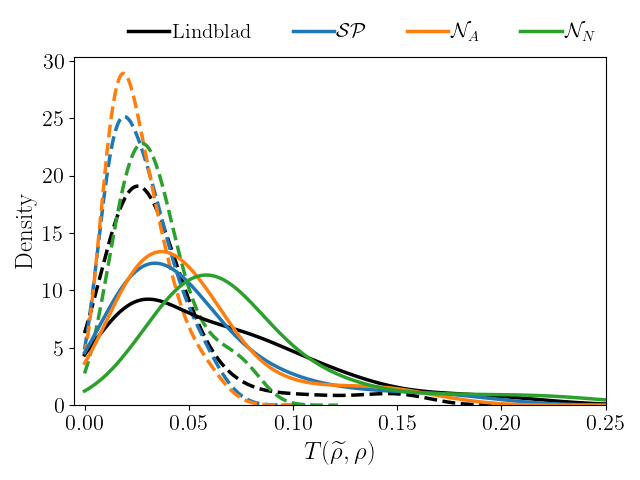} %
\label{fig:Tant:LBase:TrDist} 
\caption{Base model: Lindblad}
\end{subfigure}
\captionsetup{singlelinecheck=off,font=footnotesize}
\caption[]{Probability distribution of the trace-distance using UDE models trained with data from all experiments on \Tant ~over $10 \upmu$s (left) and $20 \upmu$s (right). Dashed lines represent the training data set while solid lines denote the validation data set.}
\label{fig:Tant:HLBase:TrDist}
\end{figure}

Figure \ref{fig:Tant:HLBase:TrDist} shows the probability distribution of the trace distance between the density matrices obtained from experiments and those obtained using the trained ansatze for different base models (top vs bottom row) and for the two different training time domains (left vs right column).
Comparing the accuracy due to size of the training dataset, the higher and more concentrated densities for low trace distances show that the accuracy of all models improves when trained over larger intervals, both for the LvN and the Lindblad base model, and across all learning models. However, we observe that the improved accuracy of the structure preserving model is marginal, suggesting that a smaller number of samples are needed to uniquely define such a UDE model. Comparing the accuracy of the different models over the training set, we observe from Figure \ref{fig:Tant:HLBase:TrDist} that all UDE models exhibit improved performance over the base models. Particularly, the structure-preserving model outperforms both the affine and nonlinear neural network models in accuracy over the training set, while the structure preserving and affine models outperform the nonlinear model in accuracy over the validation set; we observe this trend across both base models. The nonlinear neural network model, however, fails to enhance the accuracy of the already accurate Lindblad base model and only improves accuracy over the validation set when trained over the inaccurate LvN base model and on a more extended training interval.

Table \ref{table:Tant:StatsMoment} presents the mean and standard deviation of the trace distance across various base models, UDE models, and training durations. In all cases, both the structure preserving and affine neural network models consistently deliver more accurate predictions, as indicated by their low expected values, and demonstrate greater stability, as evident by their reduced variance, when compared to either of the base models. Given the relatively noise-robust nature of this QPU, the superior performance of the structure preserving and affine neural network models suggests that they are more applicable when learning the underlying quantum dynamical processes than the nonlinear model. It is worth mentioning that an alternative nonlinear model, considering different network architecture, activation functions, etc., may exhibit stronger performance for similar systems. 
\begin{table}[h]
\caption{Mean and standard deviation (in brackets) of the trace distance for different base models, UDE models, and training times using experiments on \Tant.} \label{table:Tant:StatsMoment}
\centering
\begin{tabular}{c | l | c c| c c} 
 \hline
  &  & \multicolumn{2}{c|}{Interpolation} & \multicolumn{2}{c}{Extrapolation} \\
 \hline 
& \multicolumn{1}{r| }{Training time ($\upmu$s)} &  10 & 20 & 10 & 20 \\
 \hline 
 \hline 
 Base model & UDE model  & & & & \\
 \hline 
 \hline  
 \multirow{4}{*}{\shortstack[l]{Liouville- \\ von Neumann}} & \multicolumn{1}{c|}{-} & 0.05 (0.0399)  & 0.069 (0.0375) & 0.16 (0.0615)  & 0.184 (0.0511) \\
  & Structure Preserving                                                             & 0.026 (0.0189) & 0.026 (0.015)  & 0.05 (0.0329)  & 0.048 (0.0394) \\
  & Affine                                                                           & 0.037 (0.0294) & 0.025 (0.0143) & 0.071 (0.0574) & 0.051 (0.037) \\
  & Nonlinear                                                                        & 0.045 (0.0285) & 0.037 (0.0195) & 0.177 (0.0679) & 0.113 (0.0512) \\
 \hline 
 \multirow{4}{*}{Lindblad} & \multicolumn{1}{c|}{-} & 0.037 (0.0381) & 0.037 (0.031)  & 0.057 (0.0433) & 0.065 (0.0483) \\
  & Structure Preserving                            & 0.027 (0.0172) & 0.028 (0.0153) & 0.056 (0.0336) & 0.054 (0.0418) \\
  & Affine                                          & 0.038 (0.0317) & 0.026 (0.0143) & 0.099 (0.0416) & 0.053 (0.0371) \\
  & Nonlinear                                       & 0.038 (0.0307) & 0.035 (0.0185) & 0.101 (0.0448) & 0.078 (0.0481) \\
 \hline
\end{tabular}
\end{table}

In addition to providing physically consistent time evolution of the dynamics of an open quantum system, the structure preserving ansatz allows for direct interpretability of the learned model, whereas interpretability of the affine and nonlinear models is non-trival and an active area of research. For \Tant, the learned perturbations to the Hamiltonian in the structure preserving ansatz with LvN and Lindblad base model are
\begin{align}
  \mathcal{S}^{LvN}_H = \begin{pmatrix}
    0 & 0.23-2.26i \\
    0.23+2.26i & -11.98 
    \end{pmatrix}~\mathrm{kHz}, \quad  \mathcal{S}^{Lind.}_H = \begin{pmatrix}
      0 & 0.15-2.18i \\
      0.15+2.18i & -11.32
      \end{pmatrix} ~\mathrm{kHz}
\end{align}
where both operators show a detuning perturbation of approximately 11 kHz. The off-diagonal perturbations corresponding to $\sigma_x$ and $\sigma_y$ are on the order of 0.2 KHz and 2 KHz, respectively. Note that the difference between the two learned Hamiltonian perturbations are of the order of 0.6 kHz in the diagonal and 0.08 kHz in the off-diagonals, which can be attributed to the stochasticity introduced by mini-batching during model training. The estimated decoherence times $\mathcal{T} := \cbrac{\gamma_1^{-1},\gamma_2^{-1},\gamma_3^{-1}}$ when using the LvN and Lindblad base models are $\mathcal{T}^{LvN}=\cbrac{366,366,119}~\upmu$s and $\mathcal{T}^{Lind}=\cbrac{1686,1686,688}~\upmu$s, respectively. Note that the magnitude of the dissipative perturbation are inversely proportional to the $\gamma_j^{-1}$. Hence, we see that the dissipative perturbation to the Lindblad base model is an order of magnitude smaller than it is to the LvN base model, as expected. Also, note that $\gamma_1 = \gamma_2$ since $\widecheck{\Lambda}_1 = i \widecheck{\Lambda}_2$ which corresponds to the collapse operator for energy decay. In all cases, we observed $\gamma_1 = \gamma_2$, hence, here onwards, we report only on $\gamma_1$ and $\gamma_3$. The $T_1$ decay and $T_2$ dephasing times learned by the structure preserving ansatz for the LvN base model are $183~\upmu$s and $29.8~\upmu$s, respectively. The effective $T_1$ decay and $T_2$ dephasing times (i.e. perturbed times) using the Lindblad base model are $171~\upmu$s and $27~\upmu$s, respectively. The estimated $T_2$ dephasing times are comparable to those estimates using standard characterization protocols whereas the estimated $T_1$ decay times are significantly lower.

\begin{figure}[h] \centering
\begin{subfigure}[h]{.45\textwidth} %
\centering
\includegraphicsifexists[width=\textwidth]{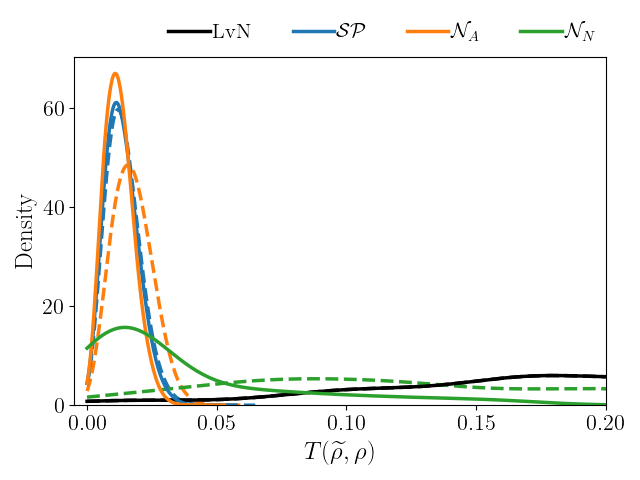} %
\label{fig:Tant:HBase:SampleId1}
\caption{Base model: Liouville von-Neumann}
\end{subfigure}
\begin{subfigure}[h]{.45\textwidth} %
\centering
\includegraphicsifexists[width=\textwidth]{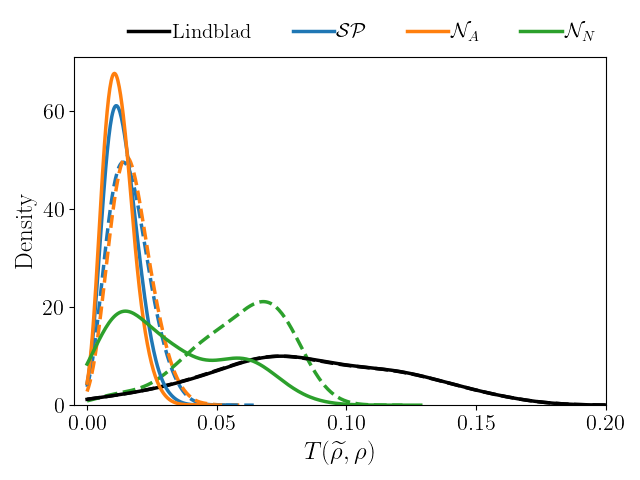} %
\label{fig:Tant:LBase:SampleId1}
\caption{Base model: Lindblad}
\end{subfigure}
\captionsetup{singlelinecheck=off,font=footnotesize}
\caption[]{Probability distribution of the trace-distance using UDE models trained over a data from a single sample on \Tant ~over a $20 \upmu$s window. Dashed and solid lines represent \emph{Experiment-Generalized} and \emph{Experiment-Specific} operators, respectively.}
\label{fig:Tant:HLBase:SampleId1}
\end{figure}

As previously mentioned, the results depicted in Fig. \ref{fig:Tant:HLBase:TrDist} were obtained by learning a single operator that minimizes the expectation of the cost function over all experiments; we will refer to such operators as \emph{Experiment-Generalized} (Exp-Gen) operators. However, in cases where the noise structure is itself experiment dependent, such an operator may not accurately capture the dynamics unique to a specific experiment. Hence, we attempt to learn operators tailored for individual experiments (i.e. $|\Vec{\xi}|=1$ in Eq. \ref{eq:OptimizationProb}); we refer to such operators as \emph{Experiment-Specific} (Exp-Spec) operators. Figure \ref{fig:Tant:HLBase:SampleId1} presents the probability density of the trace distance over a single sample using different UDE models trained using Exp-Gen and Exp-Spec approaches. It is clear that all models outperform the two base models in accuracy. We observe that the Exp-Spec operators yield more accurate predictions than the Exp-Gen counterparts, with the nonlinear model demonstrating the largest improvement. The improvement in the Exp-Spec structure preserving model over the Exp-Gen model was greatest when trained using the Lindblad base model but marginal when utilizing the LvN base model. The accuracy of the affine model using either base models was comparable. 
Table \ref{table:Tant:StatsMoment:Sample} displays the mean and standard deviation of the trace distance for the various UDE models and both base models. During interpolation, all UDE models exhibit a decrease in both mean and standard deviation by an order of magnitude when employing the LvN base model and by a factor of seven with the Lindblad equation as the base model. Similar trends were seen under extrapolation when using the structure preserving and affine models. It is worth noting that although the nonlinear model does not perform as well as the other UDE models, it does perform better (in expectation) than the base model. Furthermore, it is important to highlight that the improvement in accuracy from the generalized operator to the specialized operator is only marginal for the structure preserving model but substantial for the nonlinear model. The variance, however, remains on the same order of magnitude. This indicates that the UDE models can be made more accurate by tailoring the training dataset to the experiments of interest. 
\begin{table}[h]
\caption{Mean and standard deviation (in brackets) of the trace distance for different UDE models trained, evaluated for a single experiment on \Tant.} \label{table:Tant:StatsMoment:Sample}
\centering
 \begin{tabular}{ l | l | c c | c c} 
 \hline 
 & & \multicolumn{2}{c}{Liouville-von Neumann} & \multicolumn{2}{|c}{Lindblad} \\
 \hline 
 &  UDE model  & Interpolation &  Extrapolation & Interpolation &  Extrapolation \\ 
 \hline 
 \hline  
 &  \multicolumn{1}{c|}{-} & 0.105 (0.0475) & 0.209 (0.0304) & 0.075 (0.0475) & 0.091 (0.025) \\
 \hline 
 \multirow{3}{*}{\shortstack{\emph{Experiment} \\ \emph{Generalized}} } &   Structure Preserving & 0.012 (0.0055) & 0.015 (0.0067) & 0.013 (0.0059) & 0.018 (0.0076) \\
 &  Affine   & 0.015 (0.007) & 0.019 (0.0072) & 0.014 (0.0062) & 0.019 (0.0076) \\
 &  Nonlinear  & 0.054 (0.0255) & 0.165 (0.0494) & 0.051 (0.0225) & 0.063 (0.0133) \\
 \hline 
 \multirow{3}{*}{\shortstack{\emph{Experiment} \\ \emph{Specific}} } &   Structure Preserving & 0.011 (0.0049) & 0.015 (0.0064) & 0.011 (0.0049) & 0.015 (0.0064) \\
 &  Affine  & 0.011 (0.0047) &  0.014 (0.0057) & 0.01 (0.0047) & 0.013 (0.0058) \\
 &  Nonlinear  & 0.011 (0.0048) & 0.055 (0.045) & 0.012 (0.0051) & 0.046 (0.018) \\
 \hline
\end{tabular}
\end{table}

The learned perturbations to the Hamiltonian in the structure preserving ansatz with LvN and Lindblad base model are
\[
  \mathcal{S}^{LvN}_H = \begin{pmatrix}
    0 & 0.14-1.75i \\
    0.14+1.75i & -11.6
    \end{pmatrix}~\mathrm{kHz}, \quad  \mathcal{S}^{Lind.}_H = \begin{pmatrix}
      0 & 0.14-1.75i \\
      0.14+1.75i & -11.6
      \end{pmatrix}~\mathrm{kHz}
\]
where both base models converge to the same Hermitian perturbations. Furthermore, the perturbations are of a similar magnitude to those of the Exp-Gen operators. The estimated perturbations to decoherence times when using the LvN and Lindblad base models are $\mathcal{T}^{LvN}=\cbrac{246,141}~\upmu$s and $\mathcal{T}^{Lind}=\cbrac{580,1507}~\upmu$s, respectively. The effective $T_1$ decay times for both base models is $123~\upmu$s. The $T_2$ dephasing times using the LvN and Lindblad base models are $35.2~\upmu$s and $29.4~\upmu$s, respectively. Here, despite using different base models, the structure preserving models converged to the same Hamiltonian and similar dissipative operators.

\subsection{Learning Dynamics of QPU - \Geb} \label{subsubsec:qudit:geb}

Having demonstrated the effectiveness of the UDE approach on a more stable \Tant~QPU, we investigate its predictive accuracy when applied to the more noisy \Geb~QPU. In this analysis, we explore the effectiveness of both an \emph{Experiment-Generalized} and \emph{Experiment-Specific} operators.

\begin{figure}[h] \centering
\begin{subfigure}[h]{0.5\textwidth} %
\includegraphicsifexists[width=\textwidth]{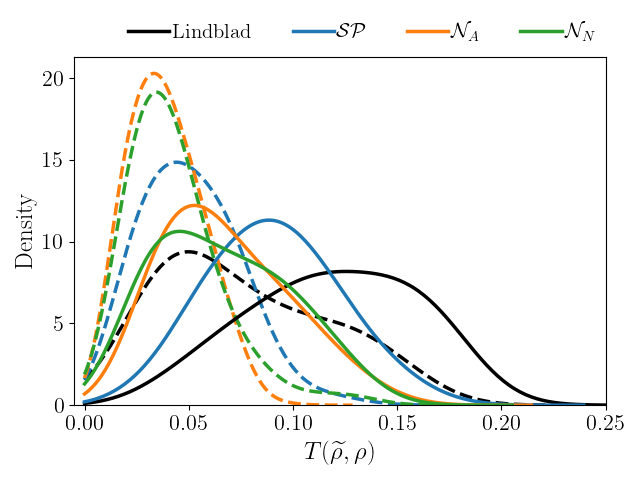}  \label{fig:Geb:TrDist} 
\end{subfigure}
\captionsetup{singlelinecheck=off,font=footnotesize}
\caption[]{Probability distribution of the trace-distance using UDE models trained over a data from all experiments on \Geb ~(Dashed and solid lines represent the measure over training and validation set, respectively).}
\label{fig:Geb:LBase:TrDist}
\end{figure}

We first apply our procedure to learn Exp-Gen UDE models using training data over a $10 \upmu$s window from five experiments (i.e. $|\Vec{\xi}|=5$ in Eq. \ref{eq:OptimizationProb}). Figure \ref{fig:Geb:LBase:TrDist} shows the probability distribution of the trace distance, and Table \ref{table:Geb:StatsMoment} summarizes the corresponding statistical moments for each model. The distribution of the trace distance shows that the learned UDE models outperform the base model in accuracy over both the training and validation data sets. Here we observe that the affine and the nonlinear neural network models exhibit superior performance when compared to the structure preserving ansatz. This suggests that the Markovian Lindblad model is unable to accurately model all processes present on the QPU, and these processes are better represented by the affine and nonlinear operators. By comparing Tab. \ref{table:Tant:StatsMoment} and Tab. \ref{table:Geb:StatsMoment}, it is apparent that the learned UDE models capture dynamics on both QPUs to similar accuracy in expectation, but variance on \Geb~is an order of magnitude higher than it is on \Tant. This is attributed to the increased levels of noise present on this QPU. Nevertheless, this demonstrates that our UDE approach is capable of effectively capturing the dynamics of noisy QPUs.
\begin{table}[h]
\caption{Mean and standard deviation (in brackets) of the trace distance for the base model and the different UDE models trained using data from \Geb ~.} \label{table:Geb:StatsMoment}
\centering
 \begin{tabular}{ l | c c } 
 \hline 
  UDE model  & Interpolation &  Extrapolation \\ 
 \hline 
 \hline  
  \multicolumn{1}{c|}{-} & 0.076 (0.0406) & 0.119 (0.0402) \\ 
   Structure Preserving  & 0.051 (0.0239) & 0.09  (0.0319) \\ 
   Affine                & 0.039 (0.0175) & 0.068 (0.0306) \\ 
   Nonlinear             & 0.043 (0.023)  & 0.066 (0.0322) \\ 
 \hline
\end{tabular}
\end{table}
The learned perturbations to the Hamiltonian in the structure preserving ansatz to the base model are
\begin{align}
   \mathcal{S}^{Lind.}_H = \begin{pmatrix}
      0 & 1.96-10.39i \\
      1.96+10.39i & -11.6
      \end{pmatrix}~\mathrm{kHz}
\end{align}
where the magnitude of detuning is similar to those obtained on \Tant. The magnitude of the perturbation for $\sigma_x$ and $\sigma_y$ is, 1.96 kHz and 10.4 kHz respectively, and are significantly larger than on \Tant. The estimated perturbations to the decoherence times are $\mathcal{T}^{Lind}=\cbrac{10,8.5}~\upmu$s. The perturbed $T_1$ decay time is approximately $4.6 \upmu$s and is an order of magnitude lower than the decay time estimated using standard characterization protocols. The effective $T_2$ dephasing time is approximately $1.6 \upmu$s and also lower than what was obtained through standard characterization.

\begin{figure}[h] \centering
\begin{subfigure}[h]{0.45\textwidth} %
\includegraphicsifexists[width=\textwidth]{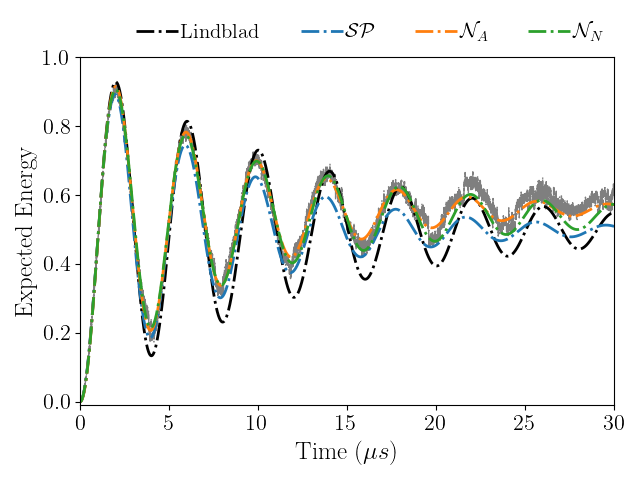}  \label{fig:Geb:Sample1:ExpEnergy} 
\end{subfigure}
\begin{subfigure}[h]{0.45\textwidth} %
\includegraphicsifexists[width=\textwidth]{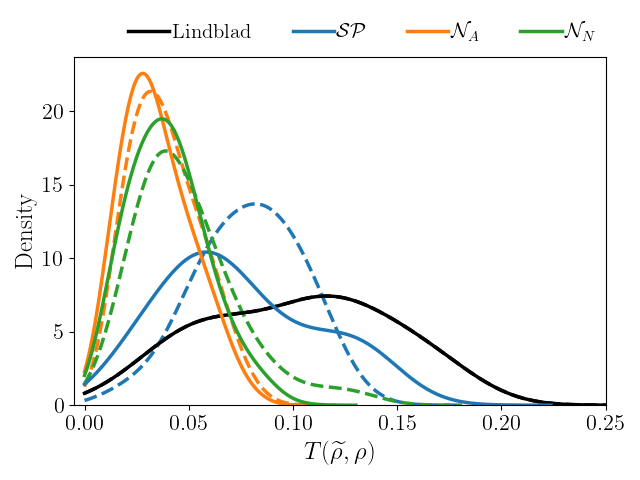}  \label{fig:Geb:Sample1:TrDist} 
\end{subfigure}
\captionsetup{singlelinecheck=off,font=footnotesize}
\caption[]{Time evolution of expected energy and probability distribution of the trace-distance using UDE models trained over data from a single sample on \Geb ~. (For the expected energy, the shaded region represent the data from the experimentally obtained density matrices; for the probability densities, dashed and solid lines represent \emph{Experiment-Generalized} and \emph{Experiment-Specific} operators, respectively).}
\label{fig:Geb:ExpEnergyTrDistOperator}
\end{figure}

Having demonstrated the UDE approach's ability to learn generalized operators, we investigate the accuracy gained by tailoring the operator for a particular experiment. Figure \ref{fig:Geb:ExpEnergyTrDistOperator} displays the expected energy, and probability density of the trace distance for a random experiment; Tab. \ref{table:Geb:StatsMoment:Sample} presents the statistical moments for the different UDE approaches when learning Exp-Spec operators. The expected energy of the density matrices computed by the affine and nonlinear models is more closely aligned with the those obtained from LIE; furthermore, the probability densities are higher and more concentrated for smaller trace distances. Although the accuracy of the Exp-Spec operators over the training data is greater than that of the Exp-Gen operators, the accuracy of the two over the validation set are comparable. The expected energy and densities of the trace distance for additional random experiments can be found in Appendix \ref{supple:geb}; these results demonstrate improved accuracy. This reinforces the conclusion that learning Exp-Spec operators greatly improves accuracy. 
\begin{table}[H]
\caption{Mean and standard deviation (in brackets) of the trace distance for different UDE models trained, evaluated for a single experiment on \Geb.} \label{table:Geb:StatsMoment:Sample}
\centering
 \begin{tabular}{ l | c c | c c} 
 \hline 
  & \multicolumn{2}{c}{\emph{Experiment-Generalized}} & \multicolumn{2}{|c}{\emph{Experiment-Specific}} \\
 \hline 
   UDE model  & Interpolation &  Extrapolation & Interpolation &  Extrapolation \\ 
 \hline 
 \hline  
   \multicolumn{1}{c|}{-} & 0.078 (0.0431) & 0.114 (0.0403) &  0.078 (0.0431) & 0.114 (0.0403) \\
 \hline 
    Structure Preserving & 0.062 (0.027) & 0.087 (0.0203) & 0.047 (0.0221) & 0.088 (0.036) \\
   Affine  & 0.031 (0.0144) & 0.042 (0.0169) & 0.024 (0.0117) & 0.04 (0.0164) \\
   Nonlinear  & 0.06 (0.0335) & 0.044 (0.0175) & 0.025 (0.0116) & 0.047 (0.0171) \\
 \hline
\end{tabular}
\end{table}

The learned perturbations to the Hamiltonian in the structure preserving ansatz to the base model are
\[
   \mathcal{S}^{Lind.}_H = \begin{pmatrix}
      0 & 1.77-6.68i \\
      1.77+6.68i & -3.74
      \end{pmatrix}~\mathrm{kHz}
\]
where the detuning introduced by the Exp-Spec operator is significantly lower than that estimated by the Exp-Gen operator. The estimated perturbed decoherence times are $\mathcal{T}^{Lind}=\cbrac{17.9,23.1}~\upmu$s, with the perturbed $T_1$ decay and $T_2$ dephasing times of approximately $7.8 \upmu$s and $2.9 \upmu$s respectively. The lower $T_1$ estimates agree with the fast decaying nature of the expected energy seen in Fig. \ref{fig:Geb:ExpEnergyTrDistOperator}, where for $N=2$, the expected energy is equivalent to the population of the first excited state.

\section{Conclusion} \label{sec:Conclusion}

This work demonstrated a data-driven approach for learning latent dynamics of quantum processing units (QPUs). The technique augments a base model of the known dynamics (e.g. Lindblad master equation) with a trained source term that models the unknown dynamics apparent in the training data from QPUs. We presented analysis with two base models: Liouville-von Neumann and Lindblad master equation, and three formulations (ansatze) for this data-driven source term: a structure preserving model, an affine model, and a nonlinear neural network model. The structure preserving ansatz offered interpretability and physically consistent time-evolution of the quantum state, whereas the affine and nonlinear models were neither directly interpretable nor yielded physically consistent evolution. The affine and nonlinear models were made physically consistent by applying a spectral filter with renormalization, \emph{a-posteriori}. The accuracy and applicability of the different ansatze was investigated using data from two different QPUs at the LLNL's Quantum Design and Integration Testbed (QuDIT), each with a different level of noise. 

We demonstrated that the structure preserving and affine models typically required less training data than the nonlinear model, while achieving improved out-of-distribution accuracy over the base models. Furthermore, the structure preserving model converged to similar operators regardless of the underlying base model used. We also showed that the structure preserving ansatz is more accurate for less-noisy QPUs where the latent dynamics are well described by the Markovian Lindblad equation, whereas the affine model performed well on both QPUs. The nonlinear model was seen to perform well only on more noisy QPUs, and hence, both the affine and nonlinear models are better able to model the noise processes. Despite the lack of interpretability and for general application of learning dynamics and noise processes, the affine model was seen to be more accurate and robust, than the remaining ansatze. Both structure preserving and affine models are applicable for learning dynamics of stable, less-noisy QPUs, and can be used as more accurate numerical models which can serve as efficient emulators of QPUs. The UDE approach identifies latent dynamics and noise processes which can greatly inform hardware development. Furthermore, accurate numerical models of the quantum devices can improve the quality of the control pulses needed for performing quantum operators accurately.

\section*{Acknowledgements}

This work was supported in part by LLNL Laboratory Directed Research and Development project 23-ERD-038 and was performed under the auspices of the U.S. Department of Energy by Lawrence Livermore National Laboratory under Contract DE-AC52-07NA27344. LLNL-JRNL-858483.

\appendix

\section{Supplementary Material: Learning Dynamics of QPU - \Geb} \label{supple:geb}

\begin{figure}[H]
\centering
\begin{subfigure}[h]{\textwidth} %
\centering   
\includegraphicsifexists[width=0.45\textwidth]{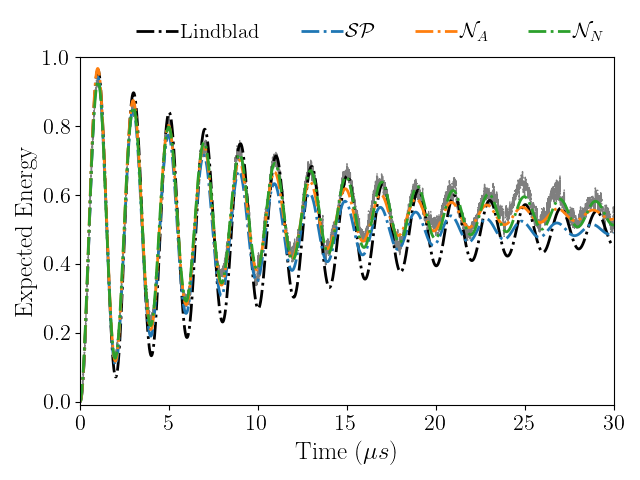}  \label{fig:Geb:Sample2:ExpEnergy} 
\includegraphicsifexists[width=0.45\textwidth]{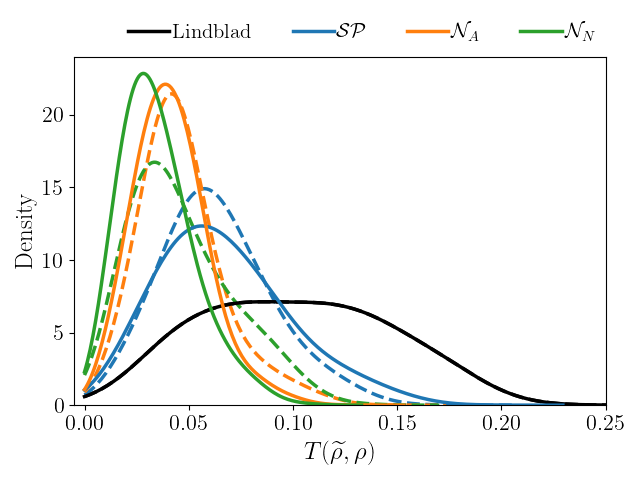}  \label{fig:Geb:Sample2:TrDist} 
\caption[]{Sample A}
\end{subfigure}
\begin{subfigure}[h]{\textwidth} %
\centering
\includegraphicsifexists[width=0.45\textwidth]{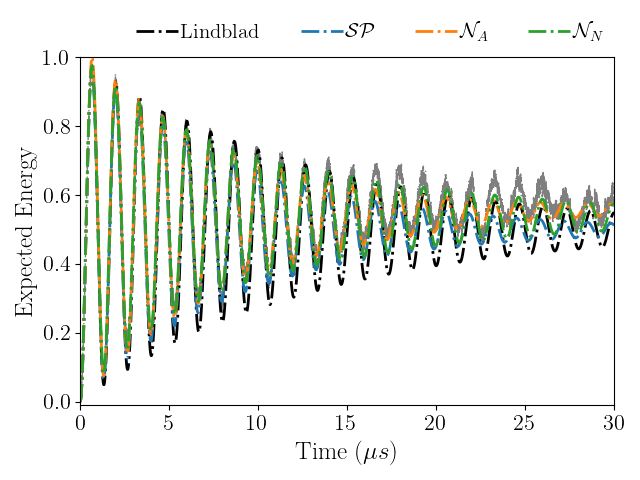}  \label{fig:Geb:Sample3:ExpEnergy} 
\includegraphicsifexists[width=0.45\textwidth]{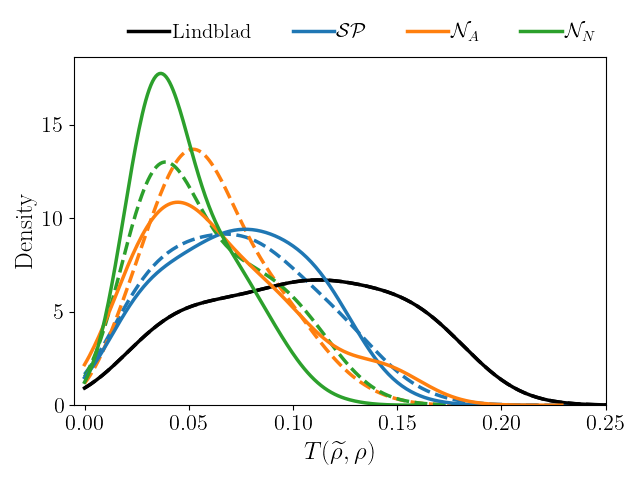}  \label{fig:Geb:Sample3:TrDist} 
\caption[]{Sample B}
\end{subfigure}
\captionsetup{singlelinecheck=off,font=footnotesize}
\caption[]{Time evolution of expected energy, probability distribution of the trace-distance using UDE models trained over a data from a single sample. Each row is a different sample (Dashed and solid lines represent \emph{Experiment-Generalized} and \emph{Experiment-Specific} operators, respectively).}
\label{fig:Geb:ExpEnergyTrDistOperator:Samples}
\end{figure}

The learned perturbations to the Hamiltonian in the structure preserving ansatz to the Lindblad base model for Sample A are
\[
   \mathcal{S}^{A}_H = \begin{pmatrix}
      0 & -2-4.01i \\
      -2+4.01i & -10.5
      \end{pmatrix} ~\mathrm{kHz}
\]
and the estimated perturbed decoherence times are $\mathcal{T}=\cbrac{11.2,11.1}~\upmu$s with the perturbed $T_1$ decay and $T_2$ dephasing times of approximately $5.1 \upmu$s and $1.9 \upmu$s respectively.

The learned perturbations to the Hamiltonian in the structure preserving ansatz to the Lindblad base model for Sample B are
\[
   \mathcal{S}^{B}_H = \begin{pmatrix}
      0 & -2.34-9.9i \\
      -2.34+9.9i & -9.57
      \end{pmatrix}~\mathrm{kHz}
\]
and the estimated perturbed decoherence times are $\mathcal{T}=\cbrac{7.8,6.2}~\upmu$s with the perturbed $T_1$ decay and $T_2$ dephasing times of approximately $3.7 \upmu$s and $1.2 \upmu$s respectively.

\bibliographystyle{unsrt}
\bibliography{references}  %

\end{document}